\documentclass[12pt]{article}

\usepackage{graphicx}
\usepackage{dcolumn}
\usepackage{bm}

\begin{document}

\title{}
\title{Theory of a Possible Mechanism for Lubrication and Surface Protection by an Electrically Neutral Hydrogels}

\author{J. B. Sokoloff\\ Department of Physics and\\Center for Interdisciplinary Research on Complex Systems\\
 Northeastern University, Boston, Massachusetts 02115, U.S.A.}

\maketitle

\begin{abstract}
\baselineskip=32pt
It is demonstrated that polymers sticking out of the surface of a neutral hydrogel are capable of preventing adhesive forces from pulling a hydrogel into close contact with a surface against which it is pressed. The proposed mechanism for lubrication or surface protection suggests a possible mechanism for protecting the cornea from a contact lens, which is held against the eye by Laplace pressure. This mechanism, however, is only able to keep a gel coated surface from sticking to a surface against which it is pressed, if the gel and surface are bathed in fluid. Expected optical properties of the gel-surface interface are discussed, in order to suggest possible ways to study the gel-solid interface experimentally. 
\end{abstract}

\baselineskip=32pt
\maketitle

\section{Introduction}

A polymer hydrogel is a cross-linked network of hydrophilic polymers\cite{1}. When a hydrogel is placed in water, the hydrophilic nature of the network chains allows it to absorb water\cite{2}. The degree of swelling is a function of various parameters, including the chemistry of the polymer, and for polyelectrolyte gels, the pH of the water\cite{2}. Because hydrogels are capable of absorbing large quantities of water, they are used in absorption applications, such as diapers and aqueous spill remediation. They also have many other applications, such as drug delivery\cite{3}, scaffolds for tissue regeneration\cite{4}, and soft contact lenses\cite{5,6,7}. Hydrogels are also popular biomimetics, because their relatively low modulus matches that of soft biological tissues\cite{8}. Polyelectrolyte hydrogels have been found to function as excellent lubricants, in the sense that the friction coefficient for the sliding of two hydrogel coated surfaces relative to each other can be very small\cite{9}. This can be explained by the fact that some of the counterions in the solution inside the hydrogels can diffuse into the interface between two hydrogels as they are pressed together. Such a thin lubricating layer of water may be what protects the cornea from abrasion when polyelectrolyte contact lenses are worn in the eye.

\begin{figure}
\center{\includegraphics [angle=0,width=6cm]{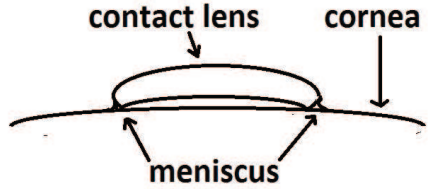}}
\caption{Illustration of a contact lens resting on the cornea.}
\label{fig4}
\end{figure}

Disposable contact lenses (one important application of hydrogels) are held in the eye by Laplace pressure that pulls the contact lens towards the cornea. See figure 1.  The Laplace pressure at the edge of the lens is given to a good approximation by
$$P=\frac{\gamma_{LV}}{R} \eqno (1)$$
where $\gamma_{LV}$ is the surface tension of the fluid and $R$ is the radius of curvature of the meniscus. 
If the meniscus is concave, $P$ is negative. (A more detailed theory of Laplace pressure for two solid surfaces placed together with fluid between them was provided by Carter\cite{12}.) The mechanism that keeps the rim of the lens and cornea from being held together by strong adhesive forces must be of microscopic origins, since there exist no macroscopic arguments to account for it. In order to prevent injury to the cornea by the lens, we need a mechanism to prevent the lens from being pulled into compete contact with the cornea. Although, as stated earlier, for lenses made from polyelectrolyte hydrogels, this can be due to counterion osmotic pressure at the interface\cite{10,11}, there must exist a mechanism that accomplishes this for neutral gels. In this article, such a mechanism will be suggested. In addition, a polymer scaling theory of the optical properties of the interface region will be presented, which will suggest features of the interface region which can be studied experimentally.

\section{Protection of Neutral Hydrogels from Adhesion by Surface Polymers} 

Many hydrogel applications, including contact lenses\cite{cont-lens} use neutral hydrogels, and hence, it is necessary to consider theoretically possible mechanisms for the maintenance of a lubricating or protective layer separating two neutral hydrogels or a neutral hydrogel in contact with a solid. We will focus here, on one possible mechanism, repulsion of the gel surfaces or a gel surface with a solid surface due to polymers sticking out from the gel surfaces, which are linked to the gel surface but not to each other, and whose mean spacing is noticeably larger than that of the polymers that make up the bulk of the gel. Such surface polymers must exist because it is unlikely that the gel surface will be perfectly smooth on the length scale of the mesh spacing of the gel. This mechanism for keeping the gel and the surface, with which it is in contact, apart is similar to that for keeping colloids from coagulating\cite{dolan}. In the mechanism described in Ref. \cite{deGennes1}, however, the monomers belonging to the polymers adsorbed on the colloid particle surfaces are assumed to be strongly attracted to the colloid particle surfaces, while in contrast, the surface polymers being considered here are the same polymers out of which the gels are constructed, and consequently, in a good solvent, they are repelled by the surface of the gel. We will now present some preliminary calculations on such surface polymers, in order to determine whether this is a viable mechanism. We will treat these polymers as flexible polymers. Neutral surface polymers are curled up, as they would be if they were in a dilute solution, as illustrated in Fig. 2a. The bulk of the gel is assumed to consist of a close packed array of attached polymer blobs\cite{32}.  It is expected that since the surface polymers described above should have spacing larger than that of the linked polymers in the bulk of the gel, these polymers do not form a polymer brush, but rather, are in the so called "mushroom regime," meaning that each polymer is curled up into a single blob. 

Then, what we have is a dilute collection of polymers in a good solvent confined between two surfaces. The free energy of a polymer compressed between two surfaces to a spacing $h$, less than the radius of gyration of a surface polymer $R_F$, is determined by the following simple argument\cite{32}: Each polymer of polymerization  (i.e., number of monomers)  $N$ breaks up into $N/g$ blobs, where $g$ is the number of monomers per blob, as illustrated in Fig. 3b. Since the radius of a blob is given by $ag^{\nu}$ \cite{32}, where for a good solvent $\nu= 3/5$ (and, for example for contact lenses,  in order for the Laplace pressure between the lens and cornea to be attractive\cite{12}, the tears must be a good solvent for the polymers in the lens). The free energy of one surface polymer when compressed between surfaces of spacing $h$ is equal to
$$k_B T\frac{N}{g}=k_B TN(\frac{a}{h})^{5/3},   \eqno (2)$$	  
since the free energy per blob of radius h is of the order of $k_B T$.  The force needed to accomplish this compression is minus the derivative of this equation with respect to $h$, or the pressure is
$$(5/3)\frac{k_B TN}{L^2}\frac{a^{5/3}}{h^{8/3}},  \eqno (3)$$	  
where $L^2$ is the surface area per polymer. Then, since the mean spacing of the points of attachment of the polymers to the gel surface is larger than the gyration radius, the total force on the polymer surface, including the van der Waals force between the two gel surfaces or between a gel and a solid surface, is
$$P=(5/3)\frac{k_B TN}{L^2}\frac{a^{5/3}}{h^{8/3}}-\frac{A_H}{6\pi h^3},   \eqno (4)$$	  
where $A_H$ is the Hameker constant. Let us set $L=\alpha^{1/2}R_F$, where $\alpha$ is the ratio of the area on the gel surface per polymer to the area occupied by one of the polymers belonging to the surface of the gel and hence it is obvious that $\alpha>1$. This expression is only valid for $h < R_F = a N^{3/5}$, and for larger $h$ it vanishes. The maximum value of P found by maximizing Eq. (4) occurs at
$$h_{max}=[\frac{9A_H}{80\pi k_B T}]^3 \alpha^3 R_F. \eqno (5)$$	       	  
The force per unit area $P$ due to the surface polymers vanishes at 
$$h=[\frac{A_h}{10\pi k_B T}]^3\alpha^3 R_F, \eqno (6)$$  
and the maximum value of P is 
$$P_{max}=\frac{5}{27}\frac{k_B T}{\alpha^9 R_F^3}[\frac{80\pi k_B T}{9A_H}]^8. \eqno (7)$$
For $h$ less than the value given in Eq. (6), the net force between the hydrogel and apposing surface is attractive. 
In order for this mechanism to produce a repulsive force, we must have $h <R_F$ but greater than the value in Eq. (6), and $\alpha$ cannot be too large.

\begin{figure}
\center{\includegraphics [angle=0,width=6cm]{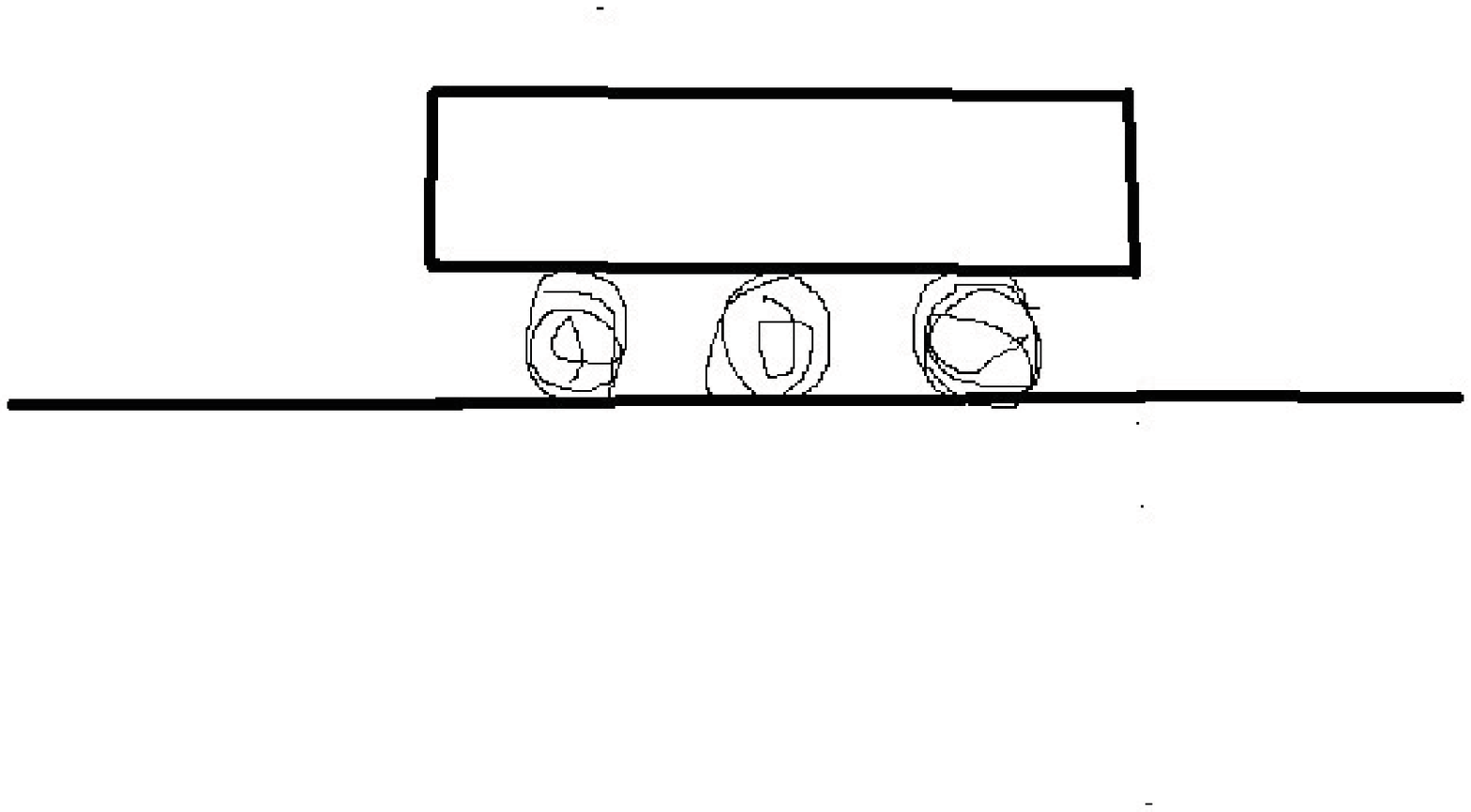} \includegraphics [angle=0,width=6cm]{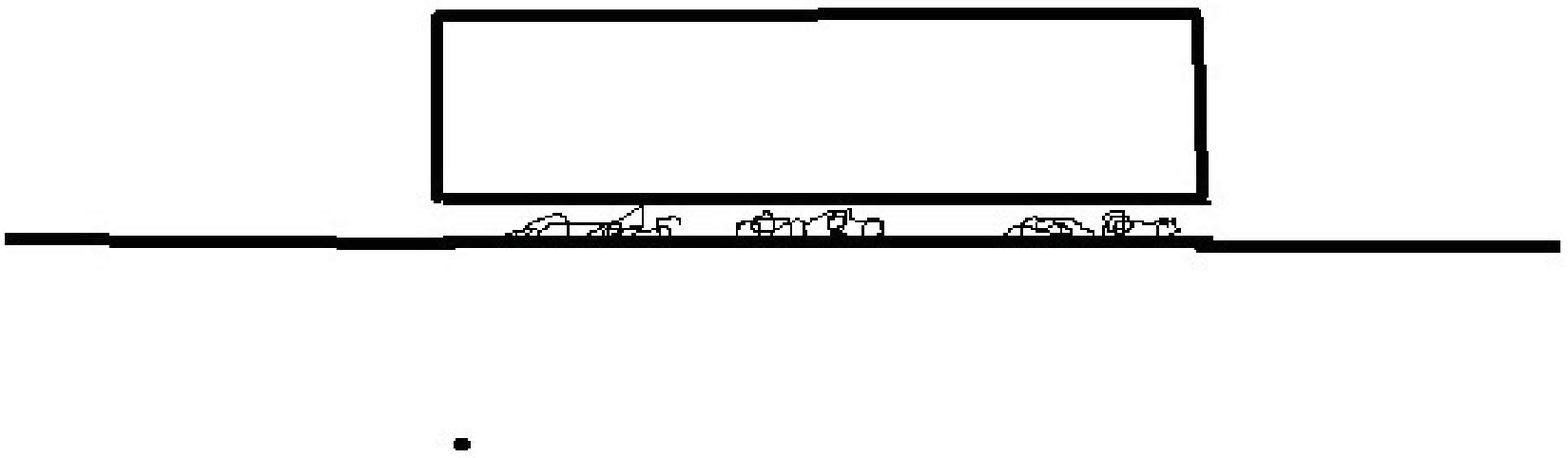}}
\caption{Illustration of a) a neutral gel and solid surface separated by surface polymers, b) compression of surface polymers as the gel and solid surface are pressed together. Here, we illustrate the break-up of surface polymers into blobs for 3 surface polymers.}
\label{fig4}
\end{figure}

The above treatment does not take into account surface roughness, however, and it is known that surface roughness can have a large effect on the van der Waals attraction\cite{33,34}. Let us consider for the moment the case of a gel in contact with a solid surface (rather than another gel surface). It is quite likely that the mean combined roughness of the solid surface and the gel surface occurs over a length scale which is large compared to length scales associated with the gel surface, such as the blob size and the polymer mesh size. Let the width of the surface roughness distribution be $b$. Then, Eq. (4) could as a first approximation be replaced by its average over a uniform distribution of interface width,  
$$P=\frac{5k_B TR_F^{5/3}}{3L^2}\frac{1}{b}\int_h^{R_F}\frac{dz}{z^{8/3}}-{1\over b}\int_h^b {A_H dz\over 6\pi z^3}=$$
$$\frac{k_B TR_F^{5/3}}{L^2 b}[\frac{1}{h^{5/3}}-\frac{1}{R_F^{5/3}}]-{A_H\over 12\pi b}[{1\over h^2}-{1\over b^2}], \eqno (8)$$
where $z$ is a mean value of the local spacing of the solid and gel surfaces and $h$ is the minimum spacing of the surfaces at which the load (or in the case of contact lenses, the Laplace pressure) is equal to the $P$. The upper limit on the first integral in Eq. (8) is $R_F$ because, as discussed earlier, the repulsive force due to surface polymer repulsion vanishes for $z>R_F$. If $b>>h$, Eq. (8) is replaced by 
$$P=\frac{k_B TR_F^{5/3}}{L^2 b}[\frac{1}{h^{5/3}}-\frac{1}{R_F^{5/3}}]-\frac{A_H}{6\pi bh^2}.   \eqno (9)$$	 
Then, $P$, the sum of the surface polymer repulsive force and the van der Waals attraction, has a maximum repulsive value of
$$P_{max}=\frac{k_B T}{\alpha R_F^2 b}[(\frac{5\pi k_B T}{A_H \alpha})^5-1] \eqno (10)$$
for 
$$h=[\frac{A_H \alpha}{5\pi k_B T}]^3 R_F.  \eqno (11)$$
Since the repulsive force due to the surface polymers is only nonzero for $h<R_F$, we require that 
$$[\frac{A_H\alpha}{5\pi k_B T}]<1.\eqno (12)$$
In order to estimate the left hand side of Eq. (12), let us now estimate $A_H$ for a gel, using the expression for it in Israelachvili's book\cite{24}, $A_H=(\pi^2/2)C\rho_1\rho_2$, where $C$ is the coefficient in the expression for the van der Waals interaction between a pair of atoms a distance $r$ apart($-C/r^6$) and $\rho_1$ and $\rho_2$ are the number densities of atoms in the gel and the surface, respectively, if the gel is pressed against a solid surface. Then $\rho_1\sim N/R_F^3=a^{3}N^{-4/5}\approx N^{-4/5}\rho_2$.  This implies that $A_H$ for a gel is equal to the product of $N^{-4/5}$ and the value of $A_H$ for a typical solid. For a contact lens on the cornea, it is likely to be even smaller because $\rho_2$ for the cornea is likely to be closer in value to the atomic number density for a gel than for a hard solid. Then, for $N=100$ and assuming that a typical value of $A_H$ for a solid is $5\times 10^{-20}J$ (which by the above argument is reduced by a factor $N^{-4/5}$), the bracketed expression in Eq. (12) is equal to $0.020\alpha$. Then, since the latter expression must be less than 1 in order for the inequality in Eq. (10) to be satisfied, we require that $\alpha$ be less than $50$. Thus, the concentration of surface polymers does not have to be too large to keep the gel from sticking to the surface with which it is in contact.

In order to be sure that the surface polymers are indeed able to prevent the gel from sticking to a surface, the van der Waals attraction between the surface polymers and the surface must be small compared to the van der Waals attraction between the bulk gel and the surface. In order to estimate the van der Waals attraction between the bulk gel and the surface with which it is in contact and the van der Waals attraction between a surface polymer and the surface, let us determine the van der Waals attraction using the methods used in Israalachvili's book\cite{24}, in particular the expression for the Hameker constant given there, $A_H=(\pi^2/2)C\rho_1\rho_2$.  When the gel and surface are not being pressed together, the attractive force between a single surface polymer and the surface is given by  $\pi^2 C\rho_1\rho_2 R_F/(6D^2)$, treating the polymer as a sphere of radius $R_F$, where $D$ is the distance between the surface of the sphere nearest to the surface and the surface. When the polymer is in contact with the surface, we account for the repulsive contribution to the force between molecules on the surface on the polymer sphere and the surface, by setting  $D\approx a$, where $a$ is a monomer size. Then the condition for the van der Waals attraction per unit area between the surface polymers and the surface with which the gel is in contact to be much smaller than the van der Waals attraction between the bulk gel and the solid surface averaged over their roughness becomes 
$$\frac{nn_p \pi^2 C\rho_1\rho_2 h}{6b}\int_a^{R_F} \frac{dz}{z^2}<<\frac{\pi C\rho_1\rho_2}{12b}\int_h^b\frac{dz}{z^3},  \eqno (13)$$ 
where $n_p=L^{-2}$ is the number of surface polymers per unit surface area and $n=N/g=(R_F/h)^{5/3}$ is the number of blobs that each surface polymer breaks up into. The upper limit on the integral on the left hand side of Eq. (13) is $R_F$ instead of $b$ because the above expression for the van der Waals interaction between two spheres must fall off rapidly for $D$ large compared to $R_F$, because we know that for $D>>R_F$ it must become proportional to $D^{-6}$, since it must reduce to the van der Waals interaction between two atoms. Eq. (13) becomes for $b>>h,R_F$
$$\frac{n_p n\pi^2 C\rho_1\rho_2 h}{6ba}<<\frac{\pi C\rho_1\rho_2}{24bh^2}.  \eqno (14)$$
The condition in Eq. (15) reduces to, using the fact that for a good solvent $g=(h/a)^{5/3}$,
$$\alpha>>\frac{4\pi N^{3/5}}{n^{4/5}}. \eqno (15)$$
 For $N=100$ and $n=20$, for example, this gives $\alpha>>18$, which can be easily satisfied, while still having $[A_H\alpha/(6\pi k_B T)]<1$, as required in order to satisfy Eq. (14). Thus, we see that it is possible for surface polymers on the surface of a neutral polymer hydrogel with properly chosen parameters to prevent adhesive contact.  

The treatment of van der Waals forces discussed here is based on Israelachvili's method of calculating them, which assumes additivity of interatomic van der Waals forces\cite{36}, which is not strictly valid for concentrated solids, but it is sufficient for getting the correct position dependence of these forces.

\section{Optical Properties of the Gel-Solid Interface}
 
Let us now describe a way to estimate the average dielectric constant of the interface from the momomer density within its volume, determined by our value for h, the surface area and our estimate for the fraction of the area occupied by the surface polymers. It is possible to use our calculations of the average dielectric constant at the interface, along with estimates of the dielectric constant inside the gel to make predictions about the reflection of light from the interface region, which we hope will  suggest a way to measure the thickness and the nature of this region optically. Before the gels (or a gel and a surface) are pressed together each surface polymer must have a density comparable to that of the bulk of the gel, since a neutral gel is to a good approximation a close packed array of polymer chains, each one of which is curled up into a blob of radius $R_F$. Since the surface polymers are not close packed, the index of refraction in the interface region should be the average of that of a surface polymer and that of the surrounding water. As the gels (or a gel and a surface) are pressed together, the polymers break up into blobs, each one of which has a higher density, and hence a higher index of refraction than the original polymer. The monomer density of the compressed surface polymers at the interface is given by $N/(\alpha R_F^2\xi)=Nn^{3/5}/ (\alpha R_F^3)$, since $\xi=ag^{3/5}=a(N/n)^{3/5}$. Since the density of the uncompressed surface polymers is given by $N/(\alpha R_F^3)$, we see that the monomer density at the interface is larger fy a factor of $n^{3/5}$. Thus, compression of the interface region is expected to increase the index of refraction there. Because $h=\xi$, we see that the monomer density of the compressed surface polymers is inversely proportional to $h$.

\section{Effects of Capillary Forces}

If the entire surface with which a neutral polymer hydrogel is in contact is coated with water, as occurs for contact lenses on the cornea of the eye, capillary forces will not be strong enough to overcome the forces due to compressed surface polymers discussed in the last section, since the radius of curvature of the meniscus will be large compared to the thickness of the interface between the rim of the contact lens (see Fig. 1) and a surface with which it is in contact. For a hydrogel with a small amount of water trapped in the interface, however, we shall see that attractive capillary forces can be larger than the repulsive force due to surface polymers. The first term in Eq. (9) gives a repulsive force per unit area
$$P=\frac{k_B TR_F^{5/3}}{L^2 b}\frac{1}{h^{5/3}}. \eqno(16)$$
For $h=R_F$, $P=k_B T/(L^2 b)=0.4Pa$ for $L=10^{-7}m$ (which corresponds to $\alpha=12.6$). The Laplace pressure for a water layer contained in the interface between the hydrogel and the surface with which it is in contact is given by Eq. (1), where $Rcos\theta=h$, where $\theta$ is the contact angle. For the purpose of getting an order of magnitude estimate, we may replace $cos\theta$ by 1. Then averaging Eq. (14) over the roughness, we obtain for the Laplace pressure
$$P_L\approx \gamma_{LV}b^{-1}\int_h^b \frac{dz}{z}=\gamma_{LV}b^{-1}ln(b/h), \eqno(17)$$
which gives for $\gamma_{LV}=7.3\times 10^{-2}$, $b=10^{-6}m$, $h=10^{-8}m$, $P_L=3.36\times 10^{7}Pa$. In order for $P$, given in Eq. (16) to be this large, we would need to have $h/R_F=1.76\times 10^{-5}$, implying $h$ smaller than a monomer, which is impossible. Thus, we conclude that neutral hydrogels can only function as lubricants or protective coatings if the interface is completely wet.

\section{Conclusions}

It was demonstrated that polymers sticking out of the surface of a neutral hydrogel are capable of preventing adhesive forces from pulling a hydrogel into close contact with a surface against which it is pressed. Possible application of this effect to non-polyelectric contact lenses is discussed.  It is predicted that if the gel and surface are pressed together with a force greater than $P_{max}$ given in Eq. (7), the adhesive force pulling them together will become larger, and a larger pull-off force will thus be required to separate them. It might be possible to test this experimentally by simply pressing different gel samples together and then separating them. For samples for which $P_{max}$ is not too large, perhaps a quartz crystal microbalance could be used to test this, for example, as has been done to study adhesion in Ref. \cite{shull}, except that a small force pressing the gel and microbalance electrode together would have to be applied. Expected optical properties of the gel-surface interface are discussed, in order to suggest possible ways to study the gel-solid interface experimentally. 

In order to obtain a complete understanding of the effects of adhesive forces, however, a multiscale treatment of them, such as that of Refs. \cite{33,34} is necessary in order to examine the conditions under which small length scale asperities can be prevented from sticking together. This will be discussed in a future publication.



\end{document}